\documentclass[12pt]{article}

\begin{document}
\begin{titlepage}
\begin{flushright}
SU-ITP-01/34 \\ hep-th/0107176 \\ July 20, 2001
\end{flushright}
\vspace{.5cm}
\begin{center}
\baselineskip=16pt {\LARGE \bf Inflation and String Cosmology\footnote{Invited talk at the conference PASCOS 2001} }\\
\vskip 2 cm

{\large Andrei
Linde} \\
\vskip 0.3cm

\

{Department of Physics, Stanford University, Stanford, CA 94305, USA
\\E-mail: alinde@stanford.edu
\\http://physics.stanford.edu/linde}
\end{center}
\vskip 2cm
\begin{center}
{\bf Abstract}
\end{center}
{Inflationary theory is already 20 years old, and it is impossible to describe all of its  versions and implications in a short talk. I will concentrate on several subjects which I believe to be most important. First of all, I will give a brief review of the first versions of inflationary theory, from Starobinsky model to new inflation. Then I will describe chaotic inflation, the theory of quantum fluctuations and density perturbations, the theory of eternal inflation, and recent observational data. In the second part of the talk I will discuss the recently proposed ekpyrotic scenario and argue that in its present form it does not provide a viable alternative to inflation.
}
\end{titlepage}
\tableofcontents{}
 
\newpage

\section{First versions of  inflationary theory}

The first   model of inflationary type was proposed by Alexei Starobinsky
in 1979~\cite{Star}. It was based on investigation of conformal anomaly
in quantum gravity. This model was rather complicated, it did not aim on
solving homogeneity, horizon and monopole problems, and it was not easy
to understand the beginning of inflation in this model. However,  it did
not suffer from the graceful exit problem, and in this sense it can be
considered the first working model of inflation. The theory of density
perturbations in this model was developed in 1981 by Mukhanov and
Chibisov~\cite{Mukh}. This was the first work where the mechanism of generation of adiabatic perturbations of metric with flat spectrum was discovered. Their theory does not differ much from  the theory
of density perturbations  in new inflation proposed   later  by
Hawking, Starobinsky, Guth, Pi, Bardeen, Steinhardt,  Turner, and
Mukhanov ~\cite{Hawk,Mukh2}.

A much simpler inflationary model with a very clear physical motivation was proposed
by Alan Guth  in 1981~\cite{Guth}.  His model, which is now called ``old
inflation,'' was based on the theory of supercooling during the
cosmological phase transitions ~\cite{Kirzhnits}. It was so attractive, and it provided such a clear explanation why inflation is necessary,
that even now all textbooks on astronomy and most of the popular books
on cosmology  describe   inflation as   exponential expansion of the
universe in a supercooled false vacuum state. It is very easy to
explain the nature of inflation in this scenario. False vacuum is a
metastable state without any fields or particles but with large  energy
density. Imagine a universe filled with such ``heavy nothing.'' When the
universe expands, empty space remains empty, so its energy density does
not change. The universe with a constant energy density expands
exponentially, thus we have inflation in the false vacuum. This process ends when the false vacuum decays due to formation of bubbles containing normal vacuum state. Bubble wall collision was supposed to make the universe hot.

Unfortunately this scenario in its original form did not work. If the probability 
of the bubble formation is large enough, they collide, and this collision made the universe very inhomogeneous. If, on the other hand, the bubble formation is exponentially suppressed, then the bubbles never collide, and the universe becomes an empty exponentially expanding space containing many separate empty bubbles.  After many attempts to overcome this problem, Guth and Weinberg~\cite{GW} and Hawking,  Moss and  Stewart~\cite{HMS}  concluded that the old   inflation scenario
cannot be improved.

The main problem of the old inflation was related to the assumption that inflation occurs in the false vacuum.  This problem was resolved with the invention of the new
inflationary theory~\cite{New}. In this theory the main stage of inflation occurs  when the inflaton field $\phi$ driving inflation  slowly rolls down to the minimum of its effective potential.  The motion of the field is of crucial importance: density perturbations produced during inflation   are inversely proportional to $\dot \phi$~\cite{Mukh,Hawk}. Thus the key difference between the new inflationary scenario and the old one is that the most important part of inflation in the new scenario does {\it not} occur in the false vacuum state.

The new inflation scenario was plagued by its own problems. This scenario
works only if the effective potential of the field $\phi$ has a very a
flat plateau near $\phi = 0$, which  is somewhat  artificial. In most
versions of this scenario the inflaton field originally could not be in a
thermal equilibrium with other matter fields. The theory of cosmological
phase transitions, which was the basis for old and new inflation, simply
did not work in such a situation. Moreover,   thermal equilibrium
requires many particles interacting with each other. This means that new
inflation could explain why our universe was so large only if it was very
large and contained many particles from the very beginning. Finally,
inflation in this theory begins very late, and during the preceding epoch
the universe could easily collapse or become so inhomogeneous that
inflation could never happen~\cite{book}. Because of all these of difficulties
no realistic versions of the new inflationary universe   scenario have
been proposed so far.

 From a more general perspective, old and new inflation represented  a
substantial but  incomplete modification of the big bang theory. It was
still assumed that the universe was in a state of thermal equilibrium
from the very beginning, that it was relatively homogeneous and large
enough to survive until the beginning of inflation, and that the stage of
inflation was just an intermediate stage of the evolution of the
universe. In the beginning of the 80's these assumptions seemed most
natural and practically unavoidable. That is why it was so difficult to
overcome a certain psychological barrier and abandon all of these
assumptions. This was done with the invention of the chaotic inflation
scenario~\cite{Chaot}. This scenario resolved all   problems of old and
new inflation. According to this scenario, inflation may occur even in
the theories with simplest potentials such as $V(\phi) \sim  \phi^n$.
Inflation may begin even if there was no thermal equilibrium in the early
universe, and it may start even close to the Planck density, in which case
the problem of initial conditions for inflation can be easily resolved~\cite{book}.

\section{Chaotic inflation}

To explain the basic idea of chaotic inflation, let us consider  the
simplest model of a scalar field $\phi$ with a mass $m$ and with the
potential energy density $V(\phi)  = {m^2\over 2} \phi^2$.
Since this function has a minimum at $\phi = 0$,  one may expect that the
scalar field $\phi$ should oscillate near this minimum. This is indeed
the case if the universe does not expand. However, one can show that in a
rapidly expanding universe  the scalar field moves down very slowly, as a
ball in a viscous liquid, viscosity being proportional to the speed of
expansion.

There are  two equations which describe evolution of a homogeneous scalar
field
 in our model, the field equation
\begin{equation}\label{1}
 \ddot\phi + 3H\dot\phi = -{V'(\phi)} \ ,
\end{equation}
and the Einstein equation
\begin{equation}\label{2}
H^2 +{k\over a^2} ={8\pi  \over 3 M_p^2}\, \left(  {1\over 2}\dot \phi
^2+V(\phi) \right) \ .
\end{equation}
Here $H = \dot a/a $ is the Hubble parameter in the universe with a scale
factor $a(t)$, $k = -1, 0, 1$ for an open, flat or closed universe
respectively, $M_p$ is the Planck mass. In the case $V = m^2\phi^2/2$,
the first equation becomes similar to the
 equation of motion for a harmonic oscillator, where instead of $x(t)$ we have
$\phi(t)$,  with a friction term $3H\dot\phi$:   .
\begin{equation}\label{1ax}
\ddot\phi + 3H\dot\phi = -m^2\phi  \ .
\end{equation}

 If   the scalar field $\phi$  initially was large,   the Hubble parameter $H$
was large too, according to the second equation. This means that the
friction term in the first equation was very large, and therefore    the
scalar field was moving   very slowly, as a ball in a viscous liquid.
Therefore at this stage the energy density of the scalar field, unlike
the  density of ordinary matter,   remained almost constant, and
expansion of the universe continued with a much greater speed than in the
old cosmological theory. Due to the rapid growth of the scale of the
universe and a slow motion of the field $\phi$, soon after the beginning
of this regime one has $\ddot\phi \ll 3H\dot\phi$, $H^2 \gg {k\over
a^2}$, $ \dot \phi ^2\ll m^2\phi^2$, so  the system of equations can be
simplified:
\begin{equation}\label{E04}
{\dot a \over a}   ={2
m\phi\over M_p}\, \sqrt { \pi  \over 3}\ , ~~~~~~  \dot\phi = -{m M_p\over 2\sqrt{3\pi}}     .
\end{equation}
The first equation shows that if the field $\phi$ changes slowly, the size of the universe in this regime
grows approximately as $e^{Ht}$, where $H = {2 m\phi\over M_p}\, \sqrt {
\pi  \over 3}\,  $. 

This is the stage of inflation, which ends when the field $\phi$ becomes much smaller than $M_p$, and the friction terms becomes small. In  realistic versions of inflationary theory the  duration of inflation could be as short as $10^{-35}$ seconds. When inflation ends, the
scalar field $\phi$ begins to   oscillate near the minimum of $V(\phi)$.
As any rapidly oscillating classical field, it looses its energy by
creating pairs of elementary particles. These particles interact with
each other and come to a state of thermal equilibrium with some
temperature $T$. From this time on, the corresponding part of the
universe can be described by the standard hot universe theory.

The main difference between inflationary theory and the old cosmology
becomes clear when one calculates the size of a typical inflationary
domain at the end of inflation. Investigation of this question    shows
that even if  the initial size of   inflationary universe  was as small
as the Plank size $l_P \sim 10^{-33}$ cm, after $10^{-35}$ seconds of
inflation   the universe acquires a huge size of   $l \sim 10^{10^{12}}$
cm!

This number is model-dependent, but in all realistic models the  size of
the universe after inflation appears to be many orders of magnitude
greater than the size of the part of the universe which we can see now,
$l \sim 10^{28}$ cm. This immediately solves most of the problems of the
old cosmological theory.

Our universe is almost exactly homogeneous on  large scale because all
inhomogeneities were stretched by a factor of $10^{10^{12}}$.  The
density of  primordial monopoles  and other undesirable ``defects''
becomes exponentially diluted by inflation.   The universe   becomes
enormously large. Even if it was a closed universe of a size
 $\sim 10^{-33}$ cm, after inflation the distance between its ``South'' and
``North'' poles becomes many orders of magnitude greater than $10^{28}$
cm. We see only a tiny part of the huge cosmic balloon. That is why
nobody  has ever seen how parallel lines cross. That is why the universe
looks so flat.

If one considers a universe which initially consisted of many domains
with chaotically distributed scalar field  $\phi$ (or if one considers
different universes with different values of the field), then  domains in
which the scalar field was too small never inflated. The main
contribution to the total volume of the universe will be given by those
domains which originally contained large scalar field $\phi$. Inflation
of such domains creates huge homogeneous islands out of initial chaos.
Each  homogeneous domain in this scenario is much greater than the size
of the observable part of the universe.

The first models of chaotic inflation were based on the theories with
polynomial potentials, such as $V(\phi) = \pm {m^2\over 2} \phi^2
+{\lambda\over 4} \phi^4$. But the main idea of this scenario is quite
generic. One should consider any particular potential $V(\phi)$,
polynomial or not, with or without spontaneous symmetry breaking, and
study all possible initial conditions without assuming that the universe
was in a state of thermal equilibrium, and that the field $\phi$ was in
the minimum of its effective potential from the very beginning
\cite{Chaot}. This scenario strongly deviated from the standard lore of
the hot big bang theory and was psychologically difficult to accept.
Therefore during the first few years after invention of chaotic inflation
many authors claimed that the idea of chaotic initial conditions is
unnatural, and made attempts to realize the new inflation scenario based
on the theory of high-temperature phase transitions, despite numerous
problems associated with it. For example, in the beginning of the 80's there were many attempts to implement new inflation in supergravity. One of the main difficulties associated with this scenario was the requirement that thermal corrections should put the inflaton field to the top of the effective potential. This requirement, which was called ``thermal constraint''~\cite{OvrutSteinhardt}, was very difficult to satisfy, which made the corresponding models very complicated. But in fact this requirement was  irrelevant because the inflaton field in these models was practically decoupled from other fields, so it was not in a state of thermal equilibrium.  Gradually  it became clear that the
idea of chaotic initial conditions is most general, and   it is much
easier to construct a consistent cosmological theory without making
unnecessary assumptions about thermal equilibrium and high temperature
phase transitions in the early universe.

Many other versions of inflationary cosmology have been proposed since
1983. Most of them are based not  on the  theory of high-temperature
phase transitions, as in old and new inflation, but on the idea of
chaotic initial conditions, which is the definitive feature of the
chaotic inflation scenario.\footnote{One should be aware of a certain terminological ambiguity. New inflation, as defined in \cite{New}, was based on two assumptions:  the potential with a very flat top at small $\phi$, and the high-temperature phase transitions that bring the field to the top of the potential.  One can have chaotic inflation in the models with a very flat top of the potential even if the new inflation scenario based on the theory of high temperature phase transitions does not work. For brevity, one may still call such models ``new inflation,'' but in fact it is a particular realization of chaotic inflation~\cite{primordial}. }

\section{\label{s7.3}Quantum fluctuations in the inflationary universe}

The vacuum structure in the  exponentially expanding universe  is much
more complicated than in ordinary Minkowski space.
 The wavelengths of all vacuum
fluctuations of the scalar field $\phi$ grow exponentially during
inflation. When the wavelength of any particular fluctuation becomes
greater than $H^{-1}$, this fluctuation stops oscillating, and its
amplitude freezes at some nonzero value $\delta\phi (x)$ because of the
large friction term $3H\dot{\phi}$ in the equation of motion of the field
$\phi$\@. The amplitude of this fluctuation then remains almost unchanged
for a very long time, whereas its wavelength grows exponentially.
Therefore, the appearance of such a frozen fluctuation is equivalent to
the appearance of a classical field $\delta\phi (x)$ that does not vanish
after averaging over macroscopic intervals of space and time.

Because the vacuum contains fluctuations of all wavelengths, inflation
leads to the continuous creation of  new perturbations of the classical
field with wavelengths greater than $H^{-1}$, i.e. with momentum $k$
smaller than $H$. One can easily understand  on dimensional grounds that the average amplitude of
 perturbations with momentum $k \sim H$ is $O(H)$. A more accurate
investigation shows that the average amplitude of perturbations generated during a time interval $H^{-1}$
(in which the universe expands by a factor of e) is given by~\cite{FordVil,book}
\begin{equation}\label{E23}
|\delta\phi(x)| \approx \frac{H}{2\pi}\ .
\end{equation}

These fluctuations in the simplest theory $m^2\phi^2/2$ give rise to adiabatic density perturbations with the wavelength
$l(cm)$ at the moment when these perturbations begin growing and the
process of the galaxy formation starts:
\begin{equation}\label{E26}
\frac{\delta \rho}{\rho} \sim   {m\over M_p}\,  \ln l(cm)\ .
\end{equation}
The definition of  ${\delta\rho\over \rho}$ used in  ~\cite{book}
corresponds to COBE data for  ${\delta\rho\over \rho} \sim 5\cdot
10^{-5}$. This gives $m \sim 10^{-6} M_p \sim 10^{13}$ GeV.

An important feature of the spectrum of density perturbations is its
flatness: $\frac{\delta \rho}{\rho}$ in our model depends on the scale
$l$ only logarithmically. For the theories with exponential potentials, the spectrum can be represented as 
\begin{equation}\label{E26a}
\frac{\delta \rho}{\rho} \sim   l^{(1-n)/2}\ .
\end{equation}
This representation is often used for the phenomenological description of various inflationary models. Exact flatness of the spectrum implies $n = 1$.  Usually $n <1$, but the models with $n > 1$ are also possible. In most of the realistic models of inflation one has $n = 1\pm 0.2$.

Flatness of the spectrum of $\frac{\delta
\rho}{\rho}$ together with flatness of the universe ($\Omega = 1$)
constitute the two most robust predictions of inflationary cosmology. It
is possible to construct models where $\frac{\delta \rho}{\rho}$ changes
in a very peculiar way, and it is also possible to construct theories
where $\Omega \not = 1$, but it is extremely difficult to do so.

\section{From the Big Bang theory to the theory of eternal inflation}

A significant step in the development of inflationary theory was the discovery of the process of
self-reproduction of inflationary universe. This process was known to
exist in old inflationary theory~\cite{Guth} and in the new one
\cite{Et1,Et2,Et3}, but it is especially surprising and leads to most profound
consequences in the context of the chaotic inflation scenario
\cite{Eternal}.

To understand the mechanism of self-reproduction one should remember that
the processes separated by distances $l$ greater than $H^{-1}$ proceed
independently of one another. This is so because during exponential
expansion the distance between any two objects separated by more than
$H^{-1}$ is growing with a speed exceeding the speed of light. As a
result, an observer in the inflationary universe can see only the
processes occurring inside the horizon of the radius  $H^{-1}$.
An important consequence of this general result is that the process of
inflation in any spatial domain of radius $H^{-1}$ occurs independently
of any events outside it. In this sense any inflationary domain of
initial radius exceeding $H^{-1}$ can be considered as a separate
mini-universe.

To investigate the behavior of such a mini-universe, with an account
taken of quantum fluctuations, let us consider an inflationary domain of
initial radius $H^{-1}$ containing sufficiently homogeneous field with
initial value $\phi \gg M_p$. Equation (\ref{E04}) implies that during a
typical time interval $\Delta t=H^{-1}$ the field inside this domain will
be reduced by $\Delta\phi = \frac{M_p^2}{4\pi\phi}$. By comparison this
expression with $|\delta\phi(x)| \approx \frac{H}{2\pi} = \sqrt{2V(\phi)
\over 3\pi M_p^2 } \sim {m\phi\over 3 M_p}$  one can easily see that if
$\phi$ is much less than $\phi^* \sim {M_p\over 3}\sqrt{M_p\over m} $,
 then the decrease of the field $\phi$
due to its classical motion is much greater than the average amplitude of
the quantum fluctuations $\delta\phi$ generated during the same time. But
for   $\phi \gg \phi^*$ one has   $\delta\phi (x) \gg \Delta\phi$.
Because the typical wavelength of the fluctuations $\delta\phi (x)$
generated during the time is $H^{-1}$, the whole domain after the time interval $\Delta t =
H^{-1}$ effectively becomes divided into $e^3 \sim 20$ separate domains
(mini-universes) of radius $H^{-1}$, each containing almost homogeneous
field $\phi - \Delta\phi+\delta\phi$.   In almost a half of these domains
the field $\phi$ grows by $|\delta\phi(x)|-\Delta\phi \approx |\delta\phi
(x)| = H/2\pi$, rather than decreases. This means that the total volume
of the universe containing {\it growing} field $\phi$ increases approximately 10 times. During the next time interval $\Delta t = H^{-1}$ the situation repeats.
Thus, after the two time  intervals $H^{-1}$ the total volume of the
universe containing the growing scalar field increases 100 times, etc.
The universe enters eternal process of self-reproduction.

Note that   $V(\phi^*) = m^2(\phi^*)^2/2 \sim m M_p^3 \ll M_p^4$, which means that eternal inflation may occur at a density much smaller than the Planck density. But it is important that eternal inflation in the context of the chaotic inflation scenario  (unlike in the new inflation) may occur also at $V(\phi) \sim M_p^4$. In this regime  the amplitude of the quantum jumps $\delta \phi\sim {H\over 2\pi}$  could be as large as $M_p$. This may lead to important consequences.

Realistic models of elementary particles involve many kinds
of scalar fields. 
The potential energy of these scalar fields may have several different
minima. This means that the same theory may have different ``vacuum
states," corresponding to different types of symmetry breaking between
fundamental interactions, and, as a result, to different laws of
low-energy physics.

One might expect that once the field is trapped by one of the minima, it should stay there. However, in our scenario this is not the case. All scalar fields can easily drift from one minimum to another if the Hubble constant $H$ is greater than the effective mass of the field. This means that during eternal inflation in the chaotic inflation scenario, where the Hubble constant can be as large as $M_p$, the fields can probe all possible minima of the effective potential (assuming that the scalar fields have masses smaller than $M_P$).

    As a result, the universe  becomes divided into infinitely many exponentially large domains that have all possible  laws of low-energy physics. Note that this
division occurs even if the whole universe originally began in the same
state, corresponding to one particular minimum of potential energy.

If this scenario is correct, then physics alone cannot provide a complete
explanation for all properties of our part of the universe.   The same
physical theory may yield large parts of the universe that have diverse
properties.  According to this scenario, we find ourselves inside a
four-dimensional domain with our kind of physical laws not because
domains with different dimensionality and with alternate properties are
impossible or improbable, but simply because our kind of life cannot
exist in other domains. Thus eternal inflation in the context of the chaotic inflation scenario provides a simple physical justification of anthropic principle~\cite{Eternal,book}.

\section{Inflation and observations}
Looking back at the development of inflationary theory, one may wonder how could it happen that  a simple  massive noninteracting scalar field $\phi$ could have such incredibly complicated dynamical properties. This field makes the universe expand, which in turn makes the motion of the field very slow, which results in inflation. The same reason that makes the motion of the field slow, leads to creation of long-wavelength fluctuations which are responsible for galaxy formation. And finally, the same mechanism that is responsible for galaxy formation may lead to eternal process of self-reproduction of the universe. 

But inflation is not just an interesting theory that can resolve many difficult problems of the standard Big Bang cosmology. Inflation made several important predictions, which can be tested by cosmological observations. Here are the most important predictions: 

1) The universe must be flat. In most models $\Omega_{total} = 1 \pm 10^{-4}$.

2) Perturbations of metric produced during inflation are adiabatic.  

3) Inflationary perturbations have flat spectrum.  In most inflationary models the spectral index  $n = 1 \pm 0.2$.

4) These perturbations are gaussian.  

5) Perturbations of metric could be scalar, vector or tensor. Inflation mostly produces scalar perturbations, but it also produces tensor perturbations with nearly flat spectrum, and it does {\it not} produce vector perturbations. There are certain relations between the properties of  scalar and tensor perturbations produced by inflation.

6) Inflationary perturbations produce specific peaks in the CMB radiation.

It is possible to violate each of these predictions if one makes this theory sufficiently complicated. For example, it is possible to produce vector perturbations of metric in the models where  cosmic strings are produced at the end of inflation, which is the case in some versions of hybrid inflation. It is possible to have open or closed inflationary universe, it is possible to have models with nongaussian isocurvature fluctuations with a non-flat spectrum. However, it is extremely difficult to do so, and most of the inflationary models satisfy the simple rules given above.  

It is not easy to test all of these predictions. The major breakthrough in this direction was achieved due to the recent measurements of the CMB anisotropy. These measurements revealed the existence of two (or perhaps even three) peaks in the CMB spectrum \cite{CMB}. Position of these peaks is consistent with predictions of the simplest inflationary models with adiabatic gaussian perturbations, with $\Omega = 1.02 \pm 0.05$, and $n = 0.96 \pm 0.1$ \cite{Lang}. This is a significant success, especially if one keeps in mind that the main competitor, the theory of topological defects and textures, was almost completely ruled out by these data.

This does not mean that all
difficulties are over and we can relax. Inflation is still
a scenario which changes with every new idea in particle theory. 
We do not know which version of inflationary theory will survive
years from now. It is absolutely clear than new observational data are
going to rule out 99\% of all inflationary models. But it does not seem
likely that they will rule out the basic idea of inflation. Inflationary
scenario is very versatile, and now, after 20 years of persistent
attempts of many physicists to propose an alternative to inflation, we
still do not know any other  way to construct a consistent cosmological
theory.  But may be we  did not try hard enough?

Since most of inflationary models are based on 4d cosmology, it would be natural to venture into the study of higher-dimensional cosmological models. In what follows we will discuss one of the recent attempts to formulate an alternative cosmological scenario. 


\section{Ekpyrotic/pyrotechnic scenario}

During the last few years there were many  attempts to construct a   
consistent   brane cosmology, see e.g.~\cite{brane,Binetruy:2000ut,DvaliTye} and references therein. One of the   
most interesting  possibilities is to use supersymmetric BPS branes in   
cosmology. Initially there was a hope that one can interpret such branes as nonsingular BPS domain wall solutions in Randall-Sundrum scenario. However, after a detailed investigation of this possibility a series of no-go theorems have been proven~\cite{KL}. The next idea was to consider singular branes. This subject appeared to be rather nontrivial. In order to study singular branes in the formalism where the bulk and brane  actions are supersymmetric it was necessary to develop a generalization of the standard supergravity formalism by including additional fields ~\cite{BKV}. In particular, to describe singular branes in 5d, it was necessary to introduce the 4-form field ${\cal A}_{\gamma\delta\epsilon\zeta}$~\cite{BKV}. This field does not have any degrees of freedom, but it plays an important role in making the theory supersymmetric.

Investigation of the BPS brane cosmology, i.e.   
the theory of interacting and moving near-BPS branes, has brought an   
additional level of complexity, both on the technical and on the   
conceptual level. 

One of the most challenging  recent  attempts  to construct 
a consistent  cosmology  based on  a picture of colliding BPS branes 
is  ekpyrotic scenario ~\cite{KOST}.
It was claimed   that the ekpyrotic scenario is based on the   
Ho\u{r}ava-Witten (HW) phenomenology~\cite{HoravaWitten},  and it solves all major   
cosmological problems without using inflation~\cite{KOST}. 
Let us describe the main idea of the ekpyrotic scenario and evaluate its claims~\cite{KKL,KKLT}.

\section{General setup for ekpyrotic universe}\label{setup}

According to the ekpyrotic scenario, our universe is described by the Ho\u{r}ava-Witten theory~\cite{HoravaWitten}. 
There is a  static  three brane solution for the space-time metric and the
dilaton $e^{\phi}$ (volume of the Calabi-Yau space) given by
\begin{eqnarray}
\nonumber & & ds^2=D(y)(-N^2d\tau^2+A^2d\vec{x}^2)+B^2 D^4(y)dy^2 \ ,
\nonumber\\ & & e^{\phi}=B D^3(y) \ , \nonumber\\ & & D(y)=\alpha y +C
\;\;\;\;\;\;\;\;\;\;\;\;\;\;\;\;\;\;\;\;\;\;\;\;\;\;{\rm
for}\;\;y<Y \\
& & \;\;\;\;\;\;\;\;\;=(\alpha-\beta)y+C+\beta Y \;\;\;\;\;\;\;{\rm
for}\;\;y>Y, \label{eq:static}
\end{eqnarray}
where $A,B,C,N$  are constants and $C>0$. The boundary branes are located
at $y=0$ and $y=R$, and the bulk brane is located at $y=Y$, where $0\leq
Y\leq R$.  The tension of the visible brane at $y=0$ is $-\alpha$ and is supposed to be {\it negative}.  The tension of the bulk brane $\beta$ is positive and the
tension of the hidden brane at $y=R$ is  positive and equals
$\alpha-\beta$. One assumes that $\beta \ll \alpha$, so the bulk brane is
relatively light. The visible brane at $y=0$ lies in the region of smaller
volume  while $y=R$ lies in the region of larger volume. Indeed, $D(0)= C$
and $D(R)= C + \alpha R$ and  $\alpha$ is positive, so $D(0) < D(R)$. This
property was considered one of the most important features of the scenario.

The light bulk brane may either appear spontaneously from the hidden brane
or it may also exist from the very beginning, i.e. one starts with two
boundary branes and one bulk brane. The three brane configuration is
assumed to be in a nearly BPS state. The
bulk brane has a kinetic term and a potential, which was added to the theory by hand. For a ``successful
example'' this potential should have a very specific form,  $V(Y)= -v e^{-m\alpha Y}$, where $m$ is some constant. Additionally  it is
assumed that at small $Y$ the potential suddenly becomes zero due to some
nonperturbative effects.

 Due to the slight contraction of the scale factor of the universe, the bulk  brane carries some residual kinetic energy
immediately before the collision with the visible brane. After the
collision, this residual kinetic energy transforms into radiation which
will be deposited in the three dimensional space of the visible  brane.
The visible brane, now filled with hot radiation, somehow begins to expand
as a flat FRW universe. 
Quantum fluctuations of the position of the bulk brane generated during
its  motion from $Y= R$ to $Y=0$  will result in  density fluctuations
with a nearly flat spectrum.  The spectrum will have a slightly blue tilt
for the exponential potential $V(Y)$. It is argued that the problems of homogeneity, isotropy, flatness and horizon do not appear it this model because the universe, according to~\cite{KOST}, initially was in a nearly BPS state, which is homogeneous.

At the first glance, the possibility to solve all major cosmological problems in the context of M-theory without use of inflation may look extremely attractive. However, the ekpyrotic scenario is rather complicated. It consists of many parts based on various assumptions. All of these parts must work together to produce the desirable result. As we will see, a closer examination reveals many problems with each of the parts of the scenario.

\section{The sign of the visible brane tension: ekpyrotic scenario versus pyrotechnic scenario}

One of the central points of the ekpyrotic scenario is that we live on a negative tension brane, and the warp factor (the volume of the Calabi-Yau space) decreases towards the visible brane.
In the original version of Ref.~\cite{KOST} one can read: {\it As we   
will see in Section VB,   it will be   necessary  for the visible brane to   
be in the small-volume region of space-time.}   The authors repeatedly emphasized that this condition is very important for their scenario and argued that it results in a distinguishing feature
of their model: a blue spectrum of density perturbations. 

However, the standard HW phenomenology~\cite{HoravaWitten} (both for standard and non-standard embedding) is based  on the assumption that the tension of the visible brane is positive, and   the warp factor increases towards the visible brane.  There were two main reasons for such an assumption. First of all, in practically all known  versions of the HW phenomenology,  with few exceptions,  a smaller group of symmetry (such as $E_6$) lives on the positive tension brane and provides the basis for GUTs, whereas the symmetry $E_8$ on the negative tension brane may remain unbroken. It is very difficult   to find  models where $E_6$ or $SU(5)$ live on the negative tension brane~\cite{Benakli:1999sy,Donagi:2001fs}.  

There is another reason why the tension of the visible brane is positive in the standard HW phenomenology~\cite{HoravaWitten}:  The square of the gauge coupling constant is inversely proportional to the Calabi-Yau volume~\cite{HoravaWitten}. On the negative tension brane this volume is greater than on the positive tension one, see e.g.~\cite{KOST}. In the standard HW phenomenology it is usually assumed that we live on the positive tension brane with small gauge coupling, ${g^2_{GUT}\over 4\pi} \sim 0.04$. On the hidden brane with negative tension the gauge coupling constant becomes large, ${g^2_{hidden}\over 4\pi} = O(1)$, which makes the gaugino condensation possible~\cite{HoravaWitten}. It is not  impossible to have a consistent phenomenology with the small gauge coupling on the hidden brane, but this is an unconventional and not well explored possibility~\cite{Benakli:1999sy}.

Thus, we believe that the ekpyrotic scenario is at odds with the standard HW phenomenology as defined in~\cite{HoravaWitten}. The relevant issue is not the standard versus non-standard embedding,  but Ho\u{r}ava-Witten phenomenology \cite{HoravaWitten} versus   
Benakli-Lalak-Pokorski-Thomas~\cite{Benakli:1999sy} phenomenology.

But why the authors of the ekpyrotic scenario have chosen  this unconventional route? First of all, their solution was based on the 2-brane solution obtained in~\cite{universe}, where the
function $D(y)$ was taken in the form $D= C + \alpha y$. However, in~\cite{universe} $\alpha$ was negative.  The authors of~\cite{KOST} used the same notation as in \cite{universe} but assumed, as equation $D= C + \alpha y$ suggests, that  $\alpha$ is positive, i.e. the brane tension  is negative,  and $D(y)$ decreases near the visible brane at $y = 0$. Another reason was related to the idea that the density perturbations are produced in their scenario because of the decrease of $D(y)$ at small $y$, which has led to the conclusion that it is  ``necessary  for the visible brane to be in the small-volume region of space-time'' ~\cite{KOST}.  

In ref.~\cite{KKL} a  simple description of generation of density   
perturbations in the ekpyrotic scenario was presented using the methods developed in the theory of   
tachyonic preheating~\cite{tach}. It was  shown in~\cite{KKL}, in   
particular, that the requirement that the visible brane must be in the   
small-volume region of space-time is not necessary. Thus there is no   
reason to abandon the standard HW phenomenology and assume that we live on   
the negative tension brane. An improved version of the ekpyrotic scenario   
based on the assumption that we live on the positive tension brane was   
called   ``pyrotechnic universe''~\cite{KKL}.\footnote{
After we have pointed out that the requirement that the visible brane tension has negative tension is not necessary~\cite{KKL}, the authors of the ekpyrotic scenario  removed the statement that the visible brane must be in the small-volume region of space-time (i.e. that it must have negative tension) from the revised version of their paper ~\cite{KOST}. They also removed  the ``justification'' of this statement in Section VB of their paper. They also removed the statement that the spectrum of perturbations in the ekpyrotic scenario must be blue.  Then they said ~\cite{Khoury:2001iy} that they never claimed that the tension of the visible brane must be negative. }  Another distinguishing feature of this scenario is that it does not make any attempts to avoid inflation in the HW theory.\footnote{Ref.~\cite{Khoury:2001iy} implies that the pyrotechnic scenario cannot work because it is based on standard embedding that does not describe bulk branes. However, this is an obvious misunderstanding: pyrotechnic scenario, just as the ekpyrotic scenario, is based on the non-standard embedding and does include the bulk brane, as one can see from Fig. 2 of Ref.~\cite{KKL}.}

Recently this situation was re-examined by Enqvist, Keski-Vakkuri and Rasanen \cite{Enqvist:2001zk}. They concluded that negative tension  on the visible brane would lead to anti-gravity. The only modification of the ekpyrotic scenario which is not ruled out by this result is the pyrotechnic scenario~\cite{KKL}.

\section{Static action and three-brane solution}

The static solution, which was the basis of the ekpyrotic scenario, was obtained in~\cite{universe} for the 2-brane configuration. It was claimed \cite{KOST,Khoury:2001iy} that the action and the solution describing the 3-brane configuration automatically follow from the 2-brane result. However, as explained in~\cite{KKLT}, it is impossible to add the third brane to the solution  using the methods developed in Ref.~\cite{universe}. To present a three-brane solution one must use the 4-form field ${\cal A}$ and the 5-form field strength ${\cal F}=d{\cal A}$ introduced in~\cite{BKV}.

An attempt to do so was made in~\cite{KOST}. However, the part of the action depending on the 5-form field strength ${\cal F}$   in~\cite{KOST},  as well as the solution for the field ${\cal F}$, was not quite correct. 

The corrected  version of the action  is given by ~\cite{KKLT}:
\begin{eqnarray}   
 & & S=\frac{M_5^3}{2}\int_{{\cal M}_5}   
d^5x\sqrt{-g}\left(R-\frac{1}{2}(\partial\phi)^2-{3\over   
2}\frac{e^{2\phi}{\cal   
F}^2}{5!}\right) \\   
& &-3\sum_{i=1}^3\alpha_iM_5^3\int_{{\cal M}_4^{(i)}}   
d^4\xi_{(i)}\Bigl(\sqrt{-h_{(i)}}e^{-\phi}   
 -\frac{\epsilon^{\mu\nu\kappa\lambda}}{4!}{\cal   
A}_{\gamma\delta\epsilon\zeta}   
\partial_{\mu}X^{\gamma}_{(i)}\partial_{\nu}X^{\delta}_{(i)}   
\partial_{\kappa}X^{\epsilon}_{(i)}\partial_{\lambda}X^{\zeta}_{(i)}\Bigr) \ ,   
\label{eq:5daction}   
\end{eqnarray}   
where $\alpha_1= -\alpha $, $\alpha_2= \alpha-\beta$, $\alpha_3=\beta$.   
The corrected form of the static solution for ${\cal A}$ is   
\begin{equation}   
{\cal A}_{0123}= + A^3 N B^{-1} D^{-1}(y)      \ ,  \ \ \ \ \ \ \ \   
             {\cal F}_{0123y}= - A^3  N B^{-1} D^{-2}(y) D'(y) \ .   
 \label{correct}   
\end{equation} 
In the action proposed in~\cite{KOST} the coefficient $3/2$ was omitted in the term $-{3\over 2}\frac{e^{2\phi}{\cal   
F}^2}{5!}$. More importantly, the solution for ${\cal   
F}$ was given as ${\cal F}_{0123y}= D^{-2}(y) D'(y)$, which was different from our solution by the coefficient $- A^3  N B^{-1}$. Note that the negative sign of this coefficient is quite significant. For the positive sign there is no cancellation between long-range forces, and the corresponding field configuration does not describe a BPS state.

\section{Origin of the potential $V(Y)$}

The solution described above is static. The brane start moving if  one adds to the bulk brane action a new term proportional to the potential $V(Y)$. This potential is supposed to appear as a result of nonperturbative effects. However, it was not demonstrated that the potential with required properties may actually emerge in the HW theory. Indeed, the potential $V(Y)$ must be very specific. It should vanish at $Y=0$, and it must be negative and behave as $- e^{-\alpha m Y}$ at large $Y$. One could expect terms like that, but in general one also obtains terms such as $\pm  e^{-\alpha m (R-Y)}$~\cite{Moore:2000fs}. Such terms, as well as power-law corrections, must be forbidden if one wants to obtain density perturbations with flat spectrum~\cite{KKL}. The only example of a calculation of the potential of such type was given in~\cite{Moore:2000fs}. In this example the terms $\pm  e^{-\alpha m (R-Y)}$ do appear, and the sum of all terms is strictly positive in the domain of validity of the approximation used in~\cite{Moore:2000fs}.

An additional important condition is that near the hidden brane the absolute value of the potential must be smaller than $e^{-120}$, because otherwise the density perturbations on the scale of the observable part of the universe will not be generated~\cite{KKL}. Also, if one adds a positive constant suppressed by the factor $\sim e^{-120}$ to $V(Y)$, inflation may begin, and density perturbations will be generated by the usual inflationary mechanism. This is something the authors of the ekpyrotic scenario are trying to avoid.

But if the nonperturbative effects responsible for $V(Y)$ are so weak, how can they compete with the strong forces which are supposed to stabilize the positions of the visible brane and the hidden brane? Until the brane stabilization mechanism is understood, it is very hard to trust any kind of ``derivation'' of the miniscule nonperturbative potential $V(Y)$ with extremely fine-tuned properties.

\section{Towards 5d cosmology}

The static BPS solution presented above is valid for branes that are not moving. It  may serve as a starting point for finding time-dependent cosmological   
solutions describing colliding branes and subsequent expansion of the universe.   However, here we encounter new problems.

The authors of the ekpyrotic scenario assumed that in order to study 5d cosmological solutions it is sufficient to take the static metric (\ref{eq:static}) and make the coefficients $A$, $N$ and $Y$ time-dependent. Instead of solving 5d equations of general relativity in 5d, they integrated the action over the 5th dimension, added by hand the term proportional to $V(Y)$, and solved the resulting 4d equations.

But 4d equations show that the overall scale factor of the universe contracts rather than expands~\cite{KOST,KKLT}. To find out whether the visible brane may expand despite the overall contraction of the universe one  needs to solve the 5d equations prior to the integration over the 5th dimension. However, the general time-dependent  metric compatible with the planar symmetry of the problem  has a much more general form than the metric ansatz of~\cite{KOST}:
\begin{equation}\label{gener}   
ds^2= -n^2(t,y)dt^2+a^2(t,y)d{\vec x}^2+b^2(t,y)dy^2 \ .   
\end{equation}   
Here the functions $a$, $b$, and $n$ depend both on $t$ and $y$, and there still is a residual freedom of transformation of the coordinates $t$ and $y$~\cite{Binetruy:2000ut}.
 
Thus there was no reason to expect that the factorized  metric ansatz used in the ekpyrotic scenario solves the 5d equations of motion. And indeed we have found that 
 the ansatz for the metric and the fields used in~\cite{KOST} does not solve the 5d equations for the theory with the additional contribution to  bulk brane action proportional to $V(Y)$~\cite{KKLT}. 
 
Of course, one may try to solve the problem using the most general metric ansatz. This is a complicated problem, but it might be possible to solve it. However, before doing so one should first reconsider the basic assumptions of this scenario. 

It might be possible to describe the motion of the brane by adding the nonperturbative brane potential to the 5d action. However, this method may not work if one wants to study the 5d geometry induced by the nonperturbative effects. 
As an illustrative example, consider two charged plates of a capacitor in   
ordinary electrodynamics, positioned at $y = 0$ and $y = Y$. If they have   
charges $q$ and $-q$, and the electric field between the plates is $E$,   
then the potential energy of the interaction between the plates can be   
represented as the ``brane potential'' $V(Y) = -qE\,Y$. However, this energy is concentrated not on the plates but in the electric field between the plates. It is possible to use the potential $V(Y)$ to study the motion of the plates. For example, if each plate has mass $M$, one can write $m\ddot Y = -V'(Y)$, just as one does for the bulk brane acceleration in the ekpyrotic scenario. But if one studies curvature of space induced by the electric field, it would be completely incorrect to replace the contribution of the electric field to the energy-momentum tensor in the bulk by the delta-functional term proportional to $V(Y)$.

Thus  one has  a lot of things to do. First of all, one needs to find a   
theory with the potential $V(Y)$ which behaves as $-e^{-\alpha m Y}$ at   
large $Y$. This potential should be   
smaller in absolute value than $e^{-120}$ near the hidden brane. Also, this potential should not receive any contributions   
proportional to $e^{-\alpha m (R-Y)}$ due to the interaction with the   
hidden brane.  One must make sure that this potential vanishes at $y = 0$, to avoid the cosmological constant problem. Then one    
should also check that the strong forces leading to the brane   
stabilization do not interfere with the extremely weak interaction   
responsible for the potential $e^{-\alpha m Y}$. One cannot ignore the   
unresolved problem of brane stabilization (which was the position taken in   
\cite{KOST}) and speculate about the inter-brane potentials suppressed by   
a factor of $e^{-120}$.   
   
When/if the theory with the desirable potential $V(Y)$ is found, one should solve equations in 5d taking into account the nonperturbative contribution to the energy-momentum tensor in the bulk.    Until the corresponding solutions are found, one has little to say about the cosmological implications of the ekpyrotic scenario.

\section{Density perturbations and the homogeneity problem}

The problems discussed above are not the only ones that remain to be solved. For example, one should find out what happens at the moment of the brane   
collision: whether the visible brane expands or collapses,  stays at the same   
place or oscillates, etc. These issues  have not been addressed in   
\cite{KOST}, and they cannot be fully analysed until the brane   
stabilization mechanism is understood.  

But let us assume for a moment that all of these problems can be solved. Will it solve all of the cosmological problems in the way inflation did? To understand it, one should first examine the mechanism of generation of density perturbations in the ekpyrotic/pyrotechnic scenario. 

As it was shown in~\cite{KKL}, this mechanism is based on the tachyonic instability with respect to generation of quantum fluctuations of the bulk brane position in the theory with the potential $V(Y) \sim e^{-\alpha m Y}$. One may represent the position of the brane $Y(x)$ as a scalar field, and find out that the long wavelength quantum fluctuations of this field grow exponentially because the effective mass squared of this field, proportional to $V''(Y)$, is negative.    A detailed theory of the development of such instabilities recently was developed in the context of the theory of tachyonic preheating~\cite{tach}.

Inhomogeneities of the brane position lead to the $x$-dependent time delay of the `big bang' (i.e. of the moment when the `brane damage' occurs and matter is created). In inflationary theory, a similar position-dependent delay of the moment of reheating leads to density perturbations~\cite{Mukh,Hawk,Mukh2}. Simple estimates based on a similar idea lead to the conclusion~\cite{KKL} that in the pyrotechnic scenario with $V(Y) \sim e^{-\alpha m Y}$ one obtains a nearly flat spectrum of perturbations with a small red tilt. (This is opposite to what was predicted in the ekpyrotic scenario.)  In the theories with potentials $\sim Y^n$ one may also obtain perturbations with a nearly flat spectrum~\cite{KKL}, but only for $|n| > 40$.\footnote{The results of our calculations were different from the ones obtained in~\cite{KOST} by a factor of $(3B)^{-1/2} \sim 20$. In the revised version of their paper~\cite{KOST}, the authors of the ekpyrotic scenario improved their result and made a significant change of the parameters of their model.}  

Recently it was argued that when one takes into account gravitational backreaction, the perturbations of the position of the bulk brane do not lead to density perturbations after the brane collision~\cite{Lyth:2001pf}. We believe that this issue is not settled yet~\cite{KKLT}. For a careful analysis of density perturbations one would need to perform a complete 5d investigation of the brane motion. This problem is not solved even for a homogeneous universe~\cite{KKLT}, so it would be premature to come to any definite conclusions with respect to density perturbations in the ekpyrotic scenario.

If density perturbations are not produced~\cite{Lyth:2001pf}, one may stop any further discussion of this scenario. Let us, however, be optimistic and assume that the theory of density perturbations developed in~\cite{KOST,KKL} is correct.

But in this case one has a new problem to consider. Indeed, tachyonic instability amplifies not only quantum perturbations, but also classical inhomogeneities \cite{KKL}. These inhomogeneities grow in the same way as the quantum fluctuations with the same wavelength. Therefore to avoid cosmological problems the initial classical inhomogeneities of the branes must be below the level of quantum fluctuations. In other words, the universe on the large scale must be ideally homogeneous from the very beginning. By evaluating the initial amplitude of quantum fluctuations on the scale corresponding to the observable part of the universe one finds that the branes must be parallel to each other with an accuracy better than
$10^{-60}$ on a scale $10^{30}$ times greater than the distance between
the branes~\cite{KKL}. 

To understand the nature of  the problem  one may compare this scenario
with inflation. Inflation removes all previously existing inhomogeneities  and simultaneously produces small density perturbations.  Meanwhile in the ekpyrotic scenario even very small initial inhomogeneities become exponentially large. Therefore instead of resolving the homogeneity problem, the ekpyrotic scenario makes this problem much worse.

A possible resolution of this problem was proposed in~\cite{KKL}. We suggested that instead of beginning in a static nearly BPS state, as in the original version of the ekpyrotic scenario, one may start with an expanding two-brane universe. If the distance between the branes were rigidly fixed, then the inhomogeneities of the branes may gradually disappear because of the cosmic expansion. The main idea was that whereas expansion can hardly suppress the relative amplitude of density perturbations ${\delta\rho\over \rho}$, it may gradually decrease the density of matter, as well as the curvature of space. Then the branes become parallel and empty, and the ekpyrotic/pyrotechnic scenario can be realized.

However,  if one considers a generic inhomogeneous regime in the early
universe, where the initial fluctuations of metric could be $O(1)$ on the
Planckian scale, and the branes were not parallel at all, then  the
non-BPS long range forces of attraction and repulsion could be dozens of
orders of magnitude greater than  $V(R)\sim 10^{-120}$. In this case we do
not see any way to make the universe even marginally homogeneous on the
scale $10^{30}$ times greater than the brane separation.

Another possibility is very similar to the one discussed above, but assumes that one starts with a single brane, which later splits into many branes \cite{Khoury:2001iy}. But we do not see how this idea could be realized in the HW theory where the sum of tensions of all branes must vanish. In any case, the possibility that the universe became uniform due to some dynamical processes differs dramatically from the original suggestion~\cite{KOST}  that the universe was homogeneous on the scale $10^{30}$ times greater than the distance between the branes because it was  in a  {\it nearly}  BPS state. (It could not be in a truly BPS state because such a state is supposed to be stable.) But why the universe should be created in a  nearly  BPS state? According to~\cite{KOST}, this is because BPS states are very special. But homogeneity is also a very special property. So why do not we simply assume that the universe must be homogeneous from the very beginning because it is a very special state? BPS state, just like a vacuum state, is a natural candidate for a {\it final} state of the universe, but we do not see why our universe must begin its evolution in this special state (or in a {\it nearly} special state).

One of the main goals of inflation was to show how one can obtain such a {\it special} state starting from a {\it general} situation where the universe could be inhomogeneous and anisotropic. But our universe is special only with an accuracy ${\delta\rho\over \rho}\sim 10^{-4}$. To explain why it is so special, the ekpyrotic scenario must explain why it was even more special  from the very beginning (almost absolutely homogeneous on an extremely large scale). This is not a resolution of the homogeneity problem, but a most complicated problem to be solved.

But let us assume for a moment that we were able to solve the homogeneity problem without using inflation. Now we still have the flatness/entropy problem to solve.   Suppose that the universe is closed, and initially it was filled with radiation with total entropy $S$. Then its total lifetime is given by $t \sim S^{2/3} M_p^{-1}$,
after which it collapses~\cite{book}. In order to survive until the
moment $t \sim k_0 \sim 10^{32} M_5 \sim 10^{34} M_p$, the universe must
have the total entropy greater than $10^{50}$.  Thus in order to explain why the total entropy (or the
total number of particles) in the observable part of the universe is
greater than $10^{88}$ one must assume that it was greater than $10^{50}$
from the very beginning. This is the so-called entropy problem~\cite{book}. If the universe initially has the Planckian temperature, its
total initial mass must be greater than $10^{50} M_p$, which is the mass problem.
Also, such a universe must have very large size from the very beginning, which is the essence of the flatness problem~\cite{book}.

In comparison, in the simplest versions of chaotic inflation scenario the
homogeneity problem is solved if our part of the universe initially was
relatively homogeneous on the smallest possible scale $O(M_p^{-1})$
\cite{Chaot}. The whole universe could have originated from a  domain with
total entropy $O(1)$ and  total mass $O(M_p)$. Once this process begins,
it leads to eternal self-reproduction of the universe in all its possible
forms~\cite{Eternal,book}. Nothing like that is possible in the ekpyrotic
scenario.

\section{Conclusions}
   
In this paper we made a brief review of the basic principles of inflationary cosmology. During the last 20 years ago this theory has considerably changed and gradually became the standard framework for the investigation of the early universe. Recent observational data brought us additional reasons to believe that we might be on the right track.  It is quite encouraging that so far the simplest versions of inflationary cosmology seem to be in a good agreement with observations. Still there are many things to do. We do not know which  version  of inflationary theory is the best.  We do not even know whether the inflaton field is a scalar field, as in old, new and chaotic inflation, or is it related to the curvature scalar, as in Starobinsky model, or is it something else, like the logarithm of the radius of compactification of the distance between the branes. It is possible to have several different stages of inflation; one could solve the homogeneity and isotropy problems, another could produce density perturbations. This latter stage may look like exponential expansion in all directions, or it may be viewed as exponential expansion on some particular hypersurface in a higher dimensional space.\footnote{A nontrivial example of such a possibility is provided by  perturbations of the wall of an expanding bubble, which behave as if they were living in de Sitter space, see e.g. \cite{VilGarr}.}   

Thus, there exist many different versions of inflationary cosmology, and many new ones will certainly appear with every new development in the theory of all fundamental interactions.  But one may wonder whether these new developments  will eventually allow us to find a consistent non-inflationary cosmological theory?  While we cannot give a general answer to this question, we hope that our investigation of the ekpyrotic scenario demonstrates  how difficult it is to construct a consistent cosmological theory without using inflation.

I am grateful to Renata Kallosh, Lev Kofman and Arkady Tseytlin for the collaboration in our analysis of the ekpyrotic/pyrotechnic scenario. This work was 
supported by NSF grant PHY-9870115 and by the   
Templeton Foundation  grant 938-COS273. 

\
   

\end{document}